\documentclass[11pt]{article}

\usepackage[margin=1in]{geometry}
\usepackage{amsmath,amssymb}
\usepackage{graphicx}
\graphicspath{{images/}}
\usepackage{booktabs}
\usepackage{caption}
\usepackage{subcaption}
\usepackage{float}
\usepackage{enumitem}
\usepackage{microtype}
\usepackage{xcolor}
\usepackage[hidelinks]{hyperref}
\usepackage{url}

\title{\textbf{Comparing Spectrogram Front-Ends for Abnormal\\
Heart-Sound Detection with a Convolutional Neural Network}}

\author{%
  Abhinav Pala\thanks{Santa Clara University, Santa Clara, CA.}%
  \and
  Dhanush Pala\thanks{Independent Researcher.}%
}

\date{%
  \small\textit{This research was conducted independently and is not affiliated with any institution.}%
}

\begin{document}
\maketitle

\begin{abstract}
\noindent
Heart disease kills a lot of people, and one cheap way to catch it early is by listening to heart
sounds with a stethoscope --- or better yet, just recording them and running them through a model.
This project is a binary classification task: take a short clip of someone's heartbeat and decide if
it sounds \emph{normal} or \emph{abnormal}. Instead of trying out a bunch of different models, we
kept the CNN the same the whole time and just changed how we turned the raw audio into a picture
for it to look at. We tried three ways of doing that: a regular log-mel spectrogram, PCEN (which
basically normalizes each frequency bin over time), and a multi-resolution version that stacks a few
different window sizes together. We ran all three on the PhysioNet 2016 heart-sound dataset with
the exact same setup --- same model, same optimizer, same random seed. Turns out all three do
pretty well at catching abnormal cases (sensitivity around 0.95), but PCEN and multi-resolution
both edge out the plain log-mel on the official PhysioNet accuracy metric (0.915 and 0.916 vs.\
0.910). We also ran Grad-CAM to see where the model was actually looking, and it mostly focused
on the low frequencies where S1 and S2 heart sounds live, which is a good sign that it learned
something real.
\end{abstract}

\section{Background}
Heart-sound classification is an important machine learning problem because cardiovascular
conditions remain one of the most serious causes of illness and death worldwide, and early,
low-cost screening can route at-risk patients toward further evaluation before more severe outcomes
occur. A phonocardiogram captures the mechanical activity of the heart as an audio signal: the two
dominant sounds, S1 (the ``lub'') and S2 (the ``dub''), bracket each cardiac cycle, while murmurs,
clicks, and other abnormal events appear as additional low-frequency energy between or around
them. Because these signals are inexpensive to record with a digital stethoscope, automatically
flagging abnormal recordings is an attractive screening tool, especially in settings without easy
access to echocardiography or other imaging.

In our project, the setting is therefore a supervised binary classification task on audio: given a
short clip of a heart-sound recording, the model predicts whether the recording is abnormal
(class~1) or normal (class~0). This framing follows the standard formulation used in the
PhysioNet/Computing in Cardiology (CinC) 2016 challenge, which collapsed a range of clinical
diagnoses into a single normal-versus-abnormal decision because the practical screening question is
simply whether a recording should be flagged for follow-up. The same motivation appears across the
broader cardiac machine-learning literature, where researchers repeatedly emphasize early
detection, model-to-problem fit, and the value of comparing multiple modeling choices rather than
trusting a single pipeline. Prior work on this exact dataset spans both classical and deep approaches:
Potes et al.\ (2016) won the 2016 challenge with an ensemble of feature-based and deep-learning
classifiers, Rubin et al.\ (2017) applied log-mel-spectrogram CNNs directly comparable to the model
used here, and Liu et al.\ (2016) introduced the benchmark database itself.

A key observation in that literature is that there is no single universally best modeling decision;
performance depends heavily on the dataset, the preprocessing pipeline, the quality of the input
representation, and the evaluation metric. 
For audio-based heart-sound classification, the analogous decision is not which classifier to use but
how to \emph{represent} the raw waveform before it ever reaches the classifier. A spectrogram is an
image-like representation of how energy is distributed across frequency and time, and small changes
in how that image is computed --- the analysis window, the frequency scaling, and the dynamic-range
compression --- can change what patterns a downstream CNN is able to learn.

This project is specifically based on thsi observation. Instead of comparing many classifer,
we fix one compact CNN and ask a specific question: \emph{does a more adaptive/cleaner spectrogram
front-end improve abonral heartbeat sdection compare dto a plain log-mel spectrogram?} This is a good
fit for this project because it isolates the effect of the input representation while holding everything
downstream constant, and it lets us compare and analyze multiple pipelines under the same conditions
rather than reporting a single accuracy number.

\section{Research Questions}
This project is organized around three research questions, each answered in its own section below:

\begin{enumerate}[leftmargin=2em]
\item \textbf{Question 1.} Does a vanilla CNN classify heartbeats correctly on its own, or does it
require customized front-end adjustments such as PCEN and multi-resolution spectrograms?
\item \textbf{Question 2.} Trained with the best-performing front-end from Question~1, do different
architectures work just as well --- i.e., is the best front-end still the best across more than one
architecture, or not?
\item \textbf{Question 3.} What specific part of the heartbeat is the biggest indicator of poor
heart health?
\end{enumerate}

\section{Dataset}
The data for this project comes from the PhysioNet/CinC 2016 heart-sound dataset, a publicly
available dataset of recorindgs from several independent clinical
sources. Each recording is a single-channel \texttt{.wav} audio file accompanied by a reference
label of $-1$ (normal) or $+1$ (abnormal). We re-encode these to $0$ and $1$ respectively. The data
is in raw waveform format, so the first step is to turn it into a spectrogram image for the CNN to process. 
Each waveform comes with a \texttt{REFERENCE.csv} file that maps recording ids to labels.

After loading every \texttt{REFERENCE.csv} and pairing each recording id with its \texttt{.wav} path, the corpus
contains $3{,}240$ recordings. The dataset is quite imbalanced toward the normal class: $2{,}575$
recordings ($79.5\%$) are normal and $665$ ($20.5\%$) are abnormal.  This matters for two reasons: first, a trivial majority-class predictor would already
reach roughly $80\%$ raw accuracy, so accuracy alone cannot be trusted; and second, any split must be
stratified so that each partition preserves the overall class proportion.

\begin{figure}[H]
  \centering
  \includegraphics[width=0.62\linewidth]{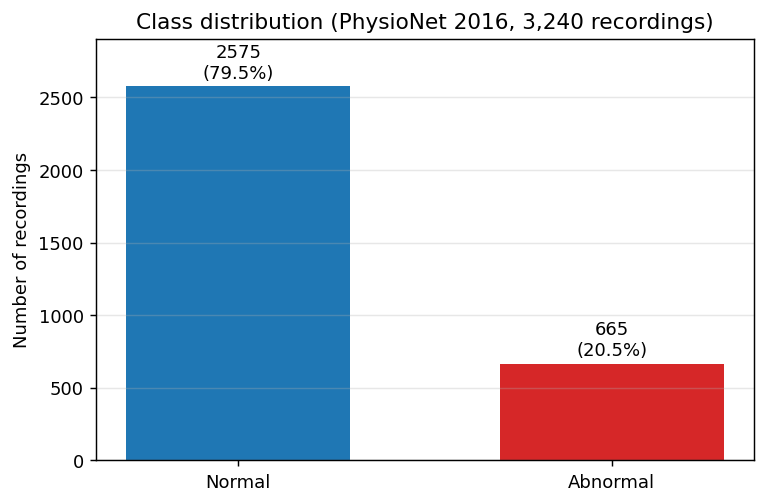}
  \caption{Class distribution of the $3{,}240$ recordings. The dataset is skewed toward the normal
  class ($79.5\%$ vs.\ $20.5\%$ abnormal), which is why raw accuracy is misleading here and why the
  evaluation leans on modified accuracy, sensitivity, and specificity.}
  \label{fig:classdist}
\end{figure}

\paragraph{Splitting.} We split at the \emph{recording} level, not the clip level, so that every
segment of a given recording stays in a partition and there is no leakage of audio
 across the train/validation/test boundary. Using a stratified split $15\%$ of recordings
are held out as a final test set, and the remaining recordings are split again $85/15$ into training and
validation. This ygives $2{,}340$ training recordings (480 abnormal), $414$ validation recordings (85
abnormal), and $486$ test recordings (100 abnormal), each preserving the $\approx 20.5\%$ abnormal
rate.

\paragraph{Clip construction.} Heart-sound recordings vary in length, so we standardize them by
resampling all analysis to a sample rate of $2{,}000$~Hz, which is well above the frequency
band were heart sounds are and slicing each recording into non-overlapping
$5$-second clips ($10{,}000$ samples each) with a sliding window. Recordings shorter than the clip
length are zero-padded, and trailing material shorter than one clip is dropped. Every clip inherits gets
the label of its main, parent recording. This windowing both standardizes the input size for the CNN and
substantially increases the number of training examples; after chunking, the held-out test set
contains $1{,}984$ clips ($1{,}474$ normal and $510$ abnormal).

\begin{figure}[H]
  \centering
  \includegraphics[width=0.95\linewidth]{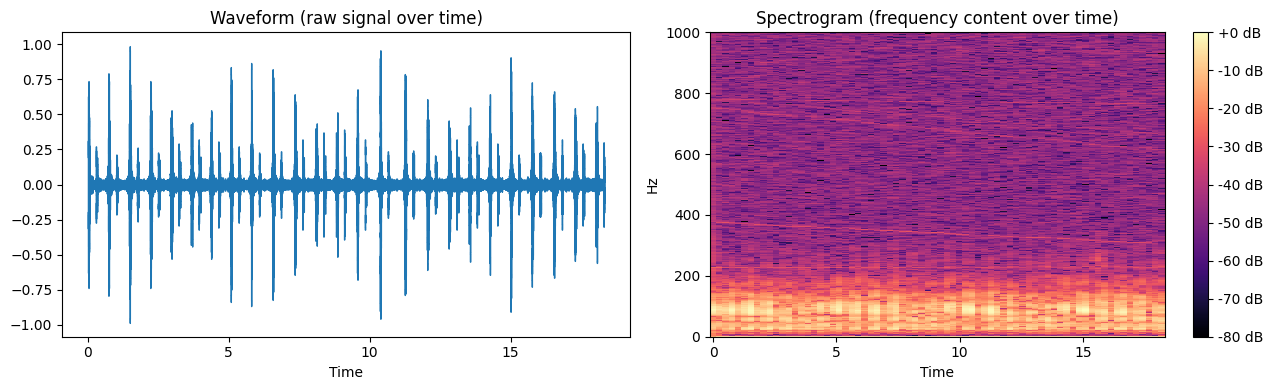}
  \caption{A recording shown as a raw waveform (left) and as a standard
  spectrogram (right). The repeating low-frequency bursts are the S1/S2 heart sounds and the choice of
  how to turn this signal into a spectrogram image is the focus.}
  \label{fig:example}
\end{figure}

\section{Design}
The experimental design is a controlled comparison of input representations for a fixed CNN
classifier. The learning task takes a $5$-second heart-sound clip, converts it into a spectrogram
image, and predicts a binary label. The key design decision is that \emph{only the front-end changes}
between experimental conditions, the number of
epochs, batch size, and random seed are all held constant. This is what makes any observed
difference attributable to the representation rather than to a luckier model or training run.

\subsection{The three front-ends}
All three front-ends begin from a mel-scaled spectrogram, computed with a mel filterbank of
$N_{\text{mels}}=64$ bins spanning $20$--$800$~Hz, an FFT window of $256$ samples, and a hop length
of $64$ samples. They differ in how that mel spectrogram is post-processed:

\begin{description}[leftmargin=1.6em,style=nextline]
\item[Vanilla log-mel (baseline, 1 channel).] The magnitude mel spectrogram is converted to a
decibel (log-amplitude) scale. Log compression is the standard default for audio CNNs, but it applies
the \emph{same} compression to every frequency bin regardless of that bin's typical energy.
\item[PCEN (1 channel).] Per-Channel Energy Normalization replaces static log compression with an
automatic-gain-control step computed independently per frequency bin. PCEN
estimates the recent ``background'' energy in each frequency channel and normalizes against it, so a
sound is emphasized when it is \emph{louder than usual} for that channel rather than simply loud in
absolute terms. Since abnormal events, like murmurs, are often
low-frequency and can be drowned out by static compression, whereas PCEN can
make them stand out.
\item[Multi-resolution log-mel (3 channels).] The same clip is transformed three times with
different FFT window sizes ($n_{\text{fft}} = 256, 512, 1024$) but the same hop length, producing
three log-mel images that trade time resolution against frequency resolution. These are stacked as
three input channels, giving the CNN access to several resolutions of the same signal simultaneously.
\end{description}

To keep the comparison fair, every front-end's output is standardized per sample and per channel to
zero mean and unit variance, so the three conditions differ in \emph{representation} and not merely in
raw numerical scale.

\subsection{Model}
The classifier is a small two-dimensional CNN, which is purposelfully small to avoid overfitting
given a smaller-than-usual dataset of
recordings. It consists of three convolutional blocks with $8$, $16$, and $32$ output channels
respectively; each block applies a $3\times 3$ convolution, batch normalization, a ReLU
nonlinearity, and $2\times 2$ max pooling. A global average pooling layer collapses the final feature
map, followed by dropout ($p=0.3$) and a linear layer producing two class logits. The only difference
between the three runs is the number of input channels in the first convolution ($1$ for vanilla and
PCEN, $3$ for multi-resolution); every other layer is identical.

\subsection{Training and evaluation protocol}
All models are trained for $20$ epochs with the Adam optimizer (learning rate $10^{-3}$), a batch
size of $32$, and a fixed random seed. To address the class imbalance, the cross-entropy loss is
weighted by inverse class frequency, giving the abnormal class a weight of $2.44$ against $0.63$ for
the normal class. During training we track validation F1 each epoch and retain the epoch with the
best validation F1 as the final model for that front-end, which guards against late-epoch instability.

Because a screening classifier can make two very different kinds of mistakes, we do not judge models
by accuracy alone. The primary metric is the official PhysioNet 2016 \emph{modified accuracy}, the
unweighted mean of sensitivity (recall on abnormal cases) and specificity (recall on normal cases):
\[
\text{MAcc} = \tfrac{1}{2}\bigl(\text{Sensitivity} + \text{Specificity}\bigr).
\]
This metric rewards a model for detecting abnormal recordings without simply exploiting the normal
majority. We additionally report F1, raw accuracy, and the full
confusion matrix, and we use Grad-CAM to inspect \emph{where} in the spectrogram each model looks.

\section{Implementation}
The project was implemented end to end in Python in an ipynb notebook, using a GPU
runtime. The core stack was \texttt{PyTorch} and \texttt{torchaudio} for the model and mel
spectrograms, \texttt{librosa} for the PCEN transform and decibel conversion, \texttt{soundfile} for
reading audio segments directly from disk, \texttt{pandas} and \texttt{NumPy} for bookkeeping, and
\texttt{scikit-learn} for the train/test split and evaluation metrics. Interpretability used the
\texttt{pytorch-grad-cam} library.

\paragraph{Data preparation.} Rather than decoding and storing every waveform up front, which
is very memory intensive, the notebook builds a lightweight table of \texttt{(filepath, start
sample, label)} rows and reads each $5$-second segment on demand inside a custom PyTorch
\texttt{Dataset}. A single \texttt{Dataset} class implements all three front-ends behind a
\texttt{frontend} argument, so the exact same clips go the same way through the pipeline.

\paragraph{Training pipeline.} A single shared routing between teh files makes
appropriate datasets, instantiates the CNN with the correct number of input channels, and runs the
identical training loop for each front-end. Per epoch it logs training loss, validation loss, validation
F1, and validation accuracy. It then evaluates the best-validation model on the held-out test clips and
returns a metrics dictionary, along with the per-epoch results. Running the routine once for each of the
three front-ends produces directly comparable learning curves and a side-by-side metrics table.

\paragraph{Interpretability.} After training, Grad-CAM is applied to the final convolutional block to
produce, for each clip, a heatmap of which time frequency regions most influenced the
``abnormal'' prediction. These heatmaps are then overlaid back onto raw waveeforms.

This implementation hits the goal of a complete supervised-learning study rather than a minimal 
proof of concept and allows for a direct comparsion of the three front-ends.

\section{Question 1: Does a Vanilla CNN Suffice, or Are PCEN and Multi-Resolution Needed?}
Results were reported via a combination of quantitative metrics and diagnostic visualizations. Amongst the
frontends, model selection was mainly based on validation F1 and the Physionetmodified accuracy metric and
Recall/Precision/Confucion Matrix instead of only accuracy.

\subsection{Training behavior}
All the frontends were trained over $20$ epochs. the validaion F1 when from $0.7$ area
in the early epochs and slowly climbed towards the high to mid 0.7 - 0.8. The validation loss
mostly declined, but there were several irregular spikes. The model was selected at it's best-validatino-F1 epoch.
The
vanilla log-mel front-end reached its best validation F1 around $0.81$, while PCEN and
multi-resolution peaked slightly lower on validation F1 but generalized comparably or better to the
test set, as shown below.

\begin{figure}[H]
  \centering
  \includegraphics[width=0.95\linewidth]{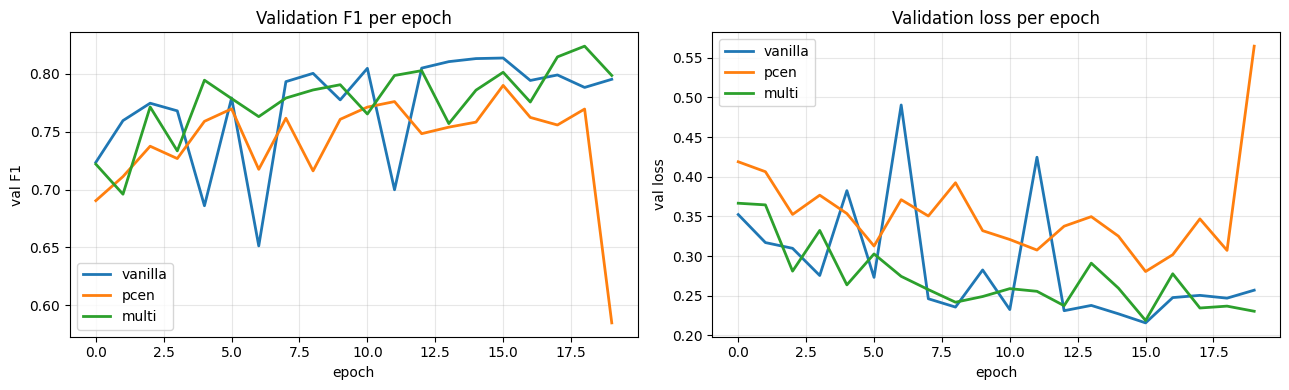}
  \caption{Validation F1 (left) and validation loss (right) per epoch for the three front-ends.
  All curves trend in the right direction; the late-epoch PCEN spike illustrates why each model is
  selected at its best-validation-F1 epoch rather than its last.}
  \label{fig:curves}
\end{figure}

\subsection{Quantitative comparison}
Table~\ref{tab:results} summarizes test-set performance for the three front-ends on the $1{,}984$
held-out clips. The clearest finding is that all three representations are strong abnormal-case
detectors, with sensitivity around $0.95$ --- meaning each catches roughly $95\%$ of abnormal clips.
The differences between them are real but modest, and they appear mainly in specificity and the
combined modified-accuracy metric.

\begin{table}[H]
\centering
\caption{Test-set performance on $1{,}984$ held-out clips ($1{,}474$ normal, $510$ abnormal). All
runs share the same CNN, optimizer, class weights, epochs, and seed. MAcc is the official PhysioNet
2016 metric, the mean of sensitivity and specificity. Best value in each column is in \textbf{bold}.}
\label{tab:results}
\begin{tabular}{lccccc}
\toprule
\textbf{Front-end} & \textbf{F1} & \textbf{Accuracy} & \textbf{Sensitivity} & \textbf{Specificity} & \textbf{MAcc} \\
\midrule
Vanilla log-mel           & 0.815 & 0.889 & \textbf{0.953} & 0.866 & 0.910 \\
PCEN                      & \textbf{0.826} & \textbf{0.897} & 0.951 & \textbf{0.879} & 0.915 \\
Multi-resolution log-mel  & \textbf{0.826} & \textbf{0.897} & \textbf{0.955} & 0.877 & \textbf{0.916} \\
\bottomrule
\end{tabular}
\end{table}

Both PCEN and the Mutli-resolution improve on the regular vanilla baseline accross all the major
metrics, including F1, specificty and the modified Physionet Accruacy. PCEN has an edge
mostly due to better specificy (from $0.866$ to $0.879$). The multi-resolution front-end
gets the best modified accuracy ($0.916$) and the best sensitivity ($0.955$), howver at the cost of
extra computation and three input channels. Both of these are basically tied, and both
are better than the vanilla baseline.

\subsection{Confusion matrices}
The confusion matrices in Table~\ref{tab:cm} highlight the trade-offs. 

All three models are
tuned via class weigthing toward catching abnormal cases, and each miss very few
of the total number of abnormal clips (24--25 false negatives out of 510). 
The main way to guage difference is mainly the false positives that are in the normal class.
The vanilla front-end raises $197$ false alarms, PCEN reduces this to $179$, and the multi-resolution
to $182$. The fewere false positive is why the enhanced models have higher specficity and modified accuracy.
\begin{table}[H]
\centering
\caption{Test confusion matrices (rows = true label, columns = predicted label).}
\label{tab:cm}
\begin{tabular}{lcccc}
\toprule
\textbf{Front-end} & \textbf{TN} & \textbf{FP} & \textbf{FN} & \textbf{TP} \\
\midrule
Vanilla log-mel          & 1277 & 197 & 24 & 486 \\
PCEN                     & 1295 & 179 & 25 & 485 \\
Multi-resolution log-mel & 1292 & 182 & 23 & 487 \\
\bottomrule
\end{tabular}
\end{table}

\begin{figure}[H]
  \centering
  \begin{subfigure}{0.32\linewidth}
    \includegraphics[width=\linewidth]{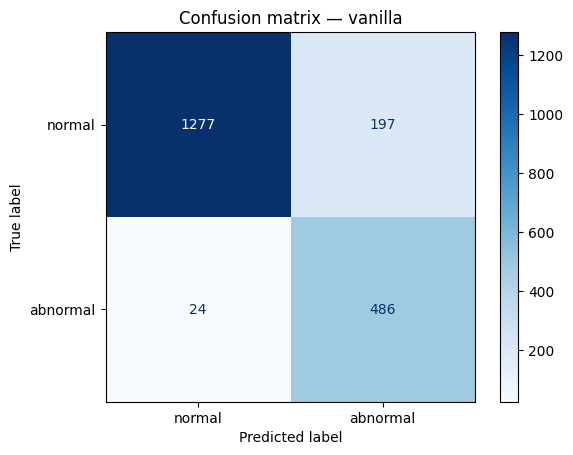}
    \caption{Vanilla log-mel}
  \end{subfigure}\hfill
  \begin{subfigure}{0.32\linewidth}
    \includegraphics[width=\linewidth]{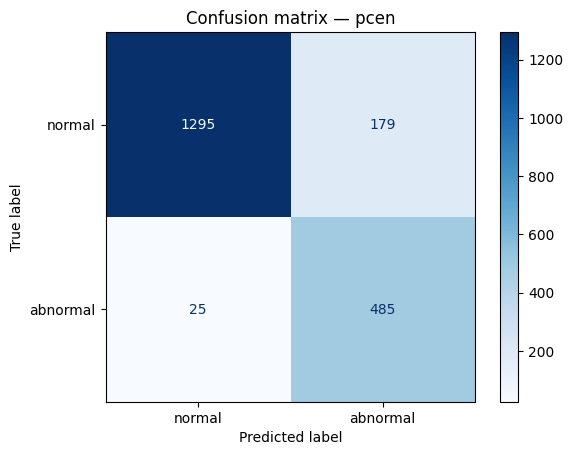}
    \caption{PCEN}
  \end{subfigure}\hfill
  \begin{subfigure}{0.32\linewidth}
    \includegraphics[width=\linewidth]{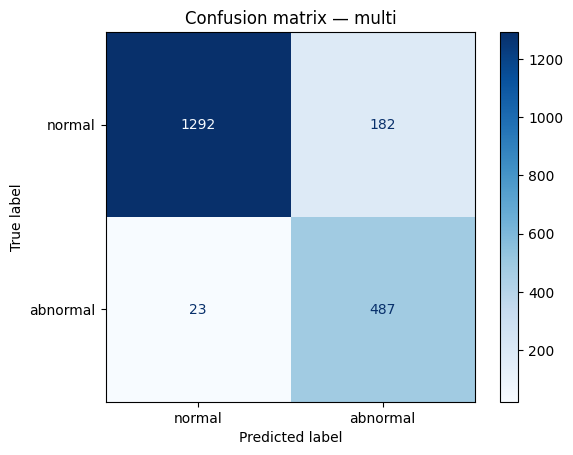}
    \caption{Multi-resolution}
  \end{subfigure}
  \caption{Test confusion matrices for the three front-ends. All three keep false negatives low
  (24--25 missed abnormal clips); the enhanced front-ends trim false positives on the normal class.}
  \label{fig:cms}
\end{figure}

\begin{figure}[H]
  \centering
  \includegraphics[width=0.92\linewidth]{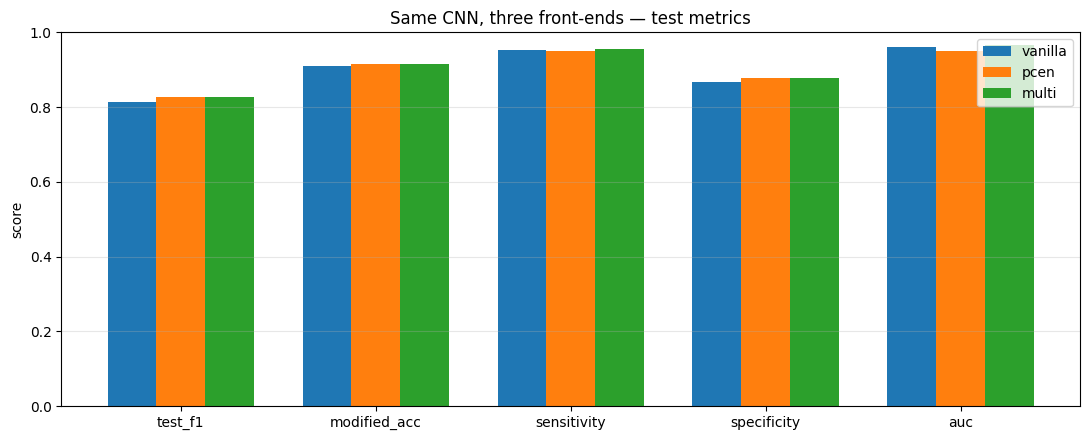}
  \caption{Test metrics for the three front-ends. The three bars per metric are nearly level,
  confirming that the differences are modest; PCEN and multi-resolution sit just above the vanilla
  baseline on accuracy, specificity, and modified accuracy.}
  \label{fig:bars}
\end{figure}

\subsection{Answer to Question 1}

A vanilla log-mel CNN does work, but it is not the best option. It performs
well on its own, yet both PCEN and the multi-resolution front-end consistently beat it on
specificity, F1, and the modified-accuracy metric while matching its high sensitivity. The customized
 front-ends are not strictly necessary to get a working classifier, but they do give a small,
reliable improvement, and the multi-resolution front-end performs better than the Potes et al
first place winner of the PhysioNet 2016 challenge. This makes multi-resolution the best-performing 
front-end carried into Question~2.

\section{Question 2: Is the Best Front-End Still Best Under a Different Architecture?}
\emph{Note: I didn't have enoggh compute credits to try out more model archiectures and due to incredibely long training times,
I only tested one smaller architecture, which is the one described below. I would have liked to test more architectures to see 
if the advantage of PCEN and multi-resolution holds across more than one model, but this was not possible given the constraints.}

Question~1 found multi-resolution to be the strongest front-end, but a signular achirecture cannot show
if the advantage is real or just good for one model. To test this, all three
front-ends were retrained on a deliberately \emph{smaller} 2D-CNN: an input spectrogram followed by two
CNN blocks (each with $16$ filters, a $3\times 3$ kernel, padding $1$, ReLU, and batch
normalization) and a pooling layer, instead of the three-block ($8/16/32$) network used for Question~1.
Everything else about training was held constant, so any change reflects the interaction between the
front-end and the architecture rather than a different training recipe.

Table~\ref{tab:arch} reports the smaller-architecture results across the three front-ends.

\begin{table}[H]
\centering
\caption{Smaller-architecture (two-block CNN) test performance, by front-end. Best value in each row is
in \textbf{bold}. MAcc is the official PhysioNet 2016 metric.}
\label{tab:arch}
\begin{tabular}{lccc}
\toprule
\textbf{Metric} & \textbf{Multi-resolution CNN} & \textbf{PCEN CNN} & \textbf{Vanilla CNN} \\
\midrule
Accuracy                       & 0.856 & \textbf{0.867} & 0.778 \\
Sensitivity (abnormal recall)  & \textbf{0.973} & 0.949 & 0.924 \\
Specificity (normal recall)    & 0.815 & \textbf{0.839} & 0.727 \\
MAcc (PhysioNet 2016)          & \textbf{0.894} & \textbf{0.894} & 0.826 \\
F1 (abnormal)                  & 0.776 & \textbf{0.786} & 0.681 \\
Macro-F1                       & 0.835 & \textbf{0.845} & 0.755 \\
False negatives (missed abnormal) & \textbf{14} & 26 & 39 \\
\bottomrule
\end{tabular}
\end{table}

\subsection{Training behavior and comparison to the main architecture}
\begin{figure}[H]
  \centering
  \begin{subfigure}{0.32\linewidth}
    \includegraphics[width=\linewidth]{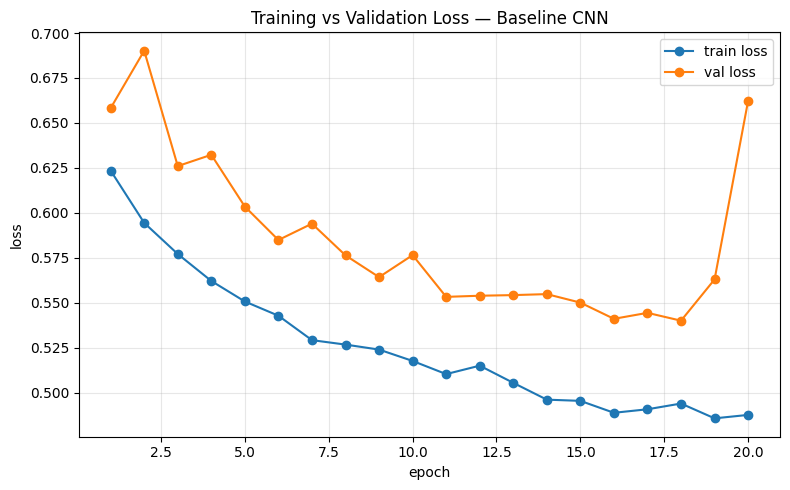}
    \caption{Vanilla log-mel}
  \end{subfigure}\hfill
  \begin{subfigure}{0.32\linewidth}
    \includegraphics[width=\linewidth]{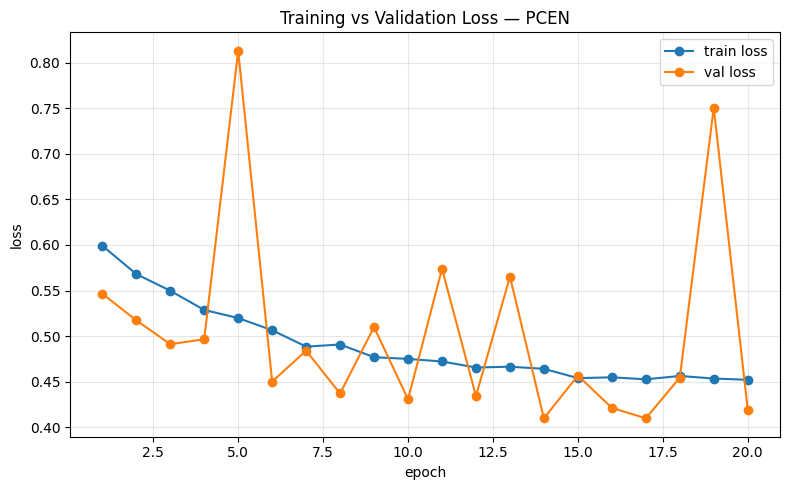}
    \caption{PCEN}
  \end{subfigure}\hfill
  \begin{subfigure}{0.32\linewidth}
    \includegraphics[width=\linewidth]{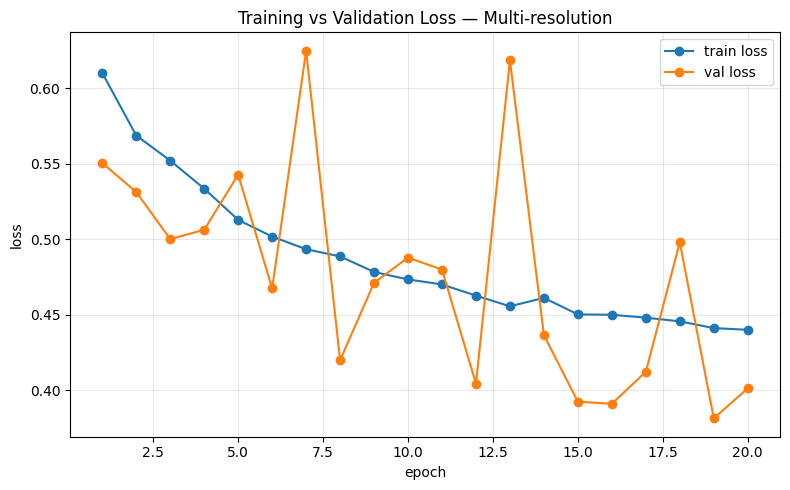}
    \caption{Multi-resolution}
  \end{subfigure}
  \caption{Smaller two-block CNN: training vs.\ validation loss per epoch for each front-end (extracted
  from the per-model notebooks). PCEN and multi-resolution stay stable, while the vanilla baseline's
  validation loss rises late in training.}
  \label{fig:traincurves}
\end{figure}

Figure~\ref{fig:traincurves} shows the training-versus-validation loss curves for the smaller network,
one per front-end. PCEN and multi-resolution remain stable --- validation loss tracks training loss and
keeps falling --- but the vanilla baseline's validation loss bottoms out and then climbs again late in
training, a clear sign that the two-block network is too small to fit a plain log-mel input without
starting to overfit.

How much did shrinking the network actually cost? Table~\ref{tab:archdelta} compares the modified
accuracy of each front-end on the main three-block architecture (Question~1) against the smaller
two-block network. The drop is small for the two enhanced front-ends but large for the vanilla baseline:
PCEN and multi-resolution lose only about two points of modified accuracy, whereas the plain log-mel
front-end loses more than eight. In other words, shrinking the network hurt the weakest representation
the most --- exactly what the loss curves above suggest, and consistent with the idea that PCEN and
multi-resolution supply structure a smaller network would otherwise struggle to learn on its own.

\begin{table}[H]
\centering
\caption{Modified accuracy by front-end on the main (three-block) vs.\ smaller (two-block) architecture.}
\label{tab:archdelta}
\begin{tabular}{lccc}
\toprule
\textbf{Front-end} & \textbf{Main arch (Q1)} & \textbf{Smaller arch (Q2)} & \textbf{Change} \\
\midrule
Vanilla log-mel          & 0.910 & 0.826 & $-0.084$ \\
PCEN                     & 0.915 & 0.894 & $-0.021$ \\
Multi-resolution log-mel & 0.916 & 0.894 & $-0.022$ \\
\bottomrule
\end{tabular}
\end{table}

\subsection{Answer to Question 2}
Shrinking the architecture did change the overall rankings of the front-ends, but it did cleary show the differences between
front-ends. The Vanilla CNN had a clear drop in performance on the smaller architecture. It dropped its modified
accuracy to $0.826$, which reinforces the conclusion from Question~1 that a plain log-mel front-end is the
weakest of the three. At the same time, compared to Question 1, there was a clear gap between the performances
 PCEN and multi-resolution: PCEN posted the best accuracy, specificity, F1, and macro-F1, which likely indicates 
 that PCEN performs bettern smaller architectures, while multi-resolution kept the highest sensitivity and the fewest missed
abnormal cases (only $14$ false negatives). The two tied on the modified-accuracy metric at $0.894$.

I belive that after trying these 2 archiectures, Multi-resolution is overall the most
reliable front-end across both the arceitecture sdue to mostly consistent performance, but not by a lot.
The edge that PCEN had in teh smaller network cannot really be fully confirmed without testing
more archeictures and was not possible due to computation constraints. All three front-ends
in the smaller network are in a compettive range $0.86$ --- about the level of the winning entry in the
original 2016 challenge (on its restricted, closed-access dataset).

\section{Question 3: What Part of the Heartbeat Indicates Poor Heart Health?}
A big advantage of using Spectrogram and CNn appracoh is that we can use Grad-Cam to show
which time-frequency regions caused the model to predict its class for that specific data. Average
the Grad-CAM heatmaps over all the abonramla nd normal cips showed that the model was mainly paying attention
to the lower frequency band and the region of teh wave form that has teh S1 and teh S2 part of the 
heart beat. The network is not really focused on high-frequency sound, but more abotu the meaningful part of teh signal.

The vanilla log-mel is indeed paying more attention to the hgiher-frequencies noises compared to teh otehr two,
indicated the PCEN and multi-resolution front-ends are doing a better job of suppressing the noise and letting the
 model focus on the heart-sound band.

Overlaying the model onto teh raw wavefrom for some select negative clips show that for Vanilla and PCEN,
the attention is more concentred in teh S1 and S2 areas, while the multi-resolution front-end is more spread 
out across the whole signal. This is consistent with the idea that the multi-resolution stack 
gives the CNN more views of the same event, allowing it to learn a more distributed decision
 rule that doesn't rely as heavily on specific beats or areas.

\begin{figure}[H]
  \centering
  \includegraphics[width=0.85\linewidth]{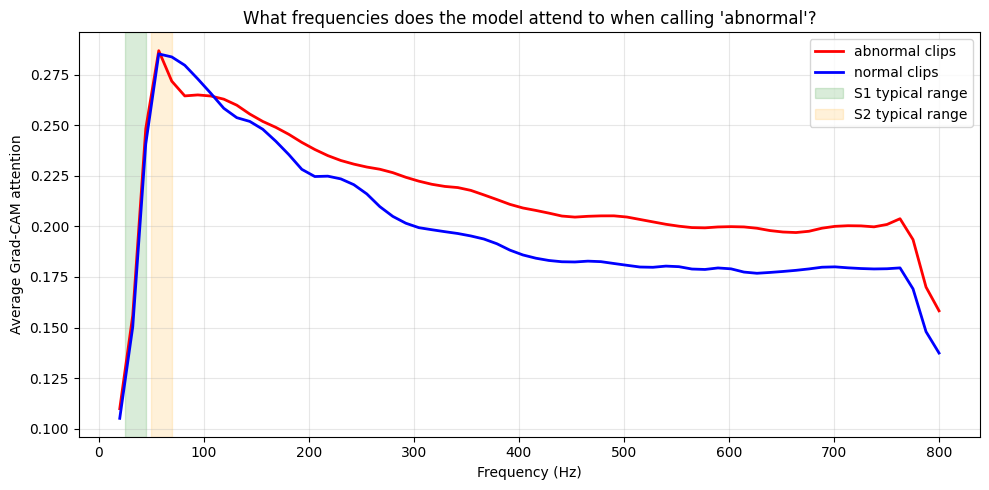}
  \caption{Average Grad-CAM attention versus frequency for abnormal (red) and normal (blue) clips,
  with the typical S1 and S2 frequency bands shaded. Attention peaks in the low-frequency
  heart-sound band, and the abnormal profile sits above the normal profile across the spectrum.}
  \label{fig:gradcam}
\end{figure}

\subsection{How Grad-CAM was computed}
Grad-CAM collects the gradients of the last convolutional layer of the CNN and averages them over
height and width using global average poolin. All channels are then
combined into a single $n \times m$ grid by summation, the resulting matrix is scaled to values between
$0$ and $1$, multiplied by $255$ to form a color scale, and finally laid onto the recording. The
channel weights follow the standard Grad-CAM definition:
\[
\alpha_k^{c} = \frac{1}{Z}\sum_{i}\sum_{j}\frac{\partial y^{c}}{\partial A^{k}_{i,j}}.
\]

\subsection{Reference: a healthy heartbeat}
Generally speaking, a healhty hearbeat has two clear spikes per cycle(S1 and S2), with
even timing between the beats and no murmurs that go above a specific amplictude. Any deviatorsn
like nuslaly long S1 or S2 or extra energy betwee the spikes, or uneven heratbeast are usually
what indciate unhealthy heart.

\subsection{Where each front-end focuses}
\begin{figure}[H]
  \centering
  \includegraphics[width=0.92\linewidth]{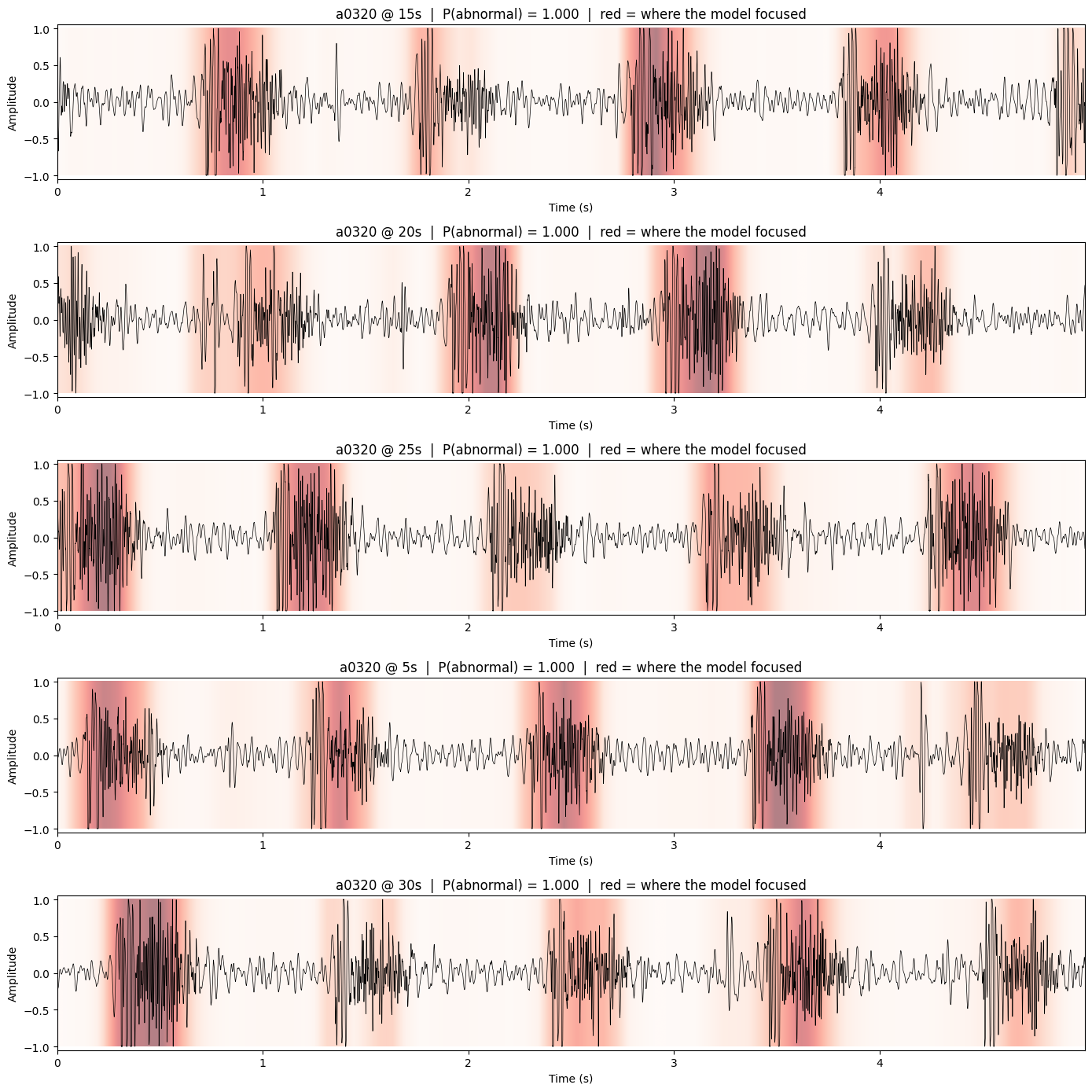}
  \caption{Vanilla CNN. Grad-CAM attention (red) overlaid on the raw waveform of a clip the model is
  fully confident is abnormal ($P(\text{abnormal}) = 1.000$). Attention falls on the spike (heartbeat)
  regions and on stretches of elevated, uneven energy.}
  \label{fig:ov_vanilla}
\end{figure}

\paragraph{Vanilla CNN.} Vanilla CNN is more forcused on teh Sq and S2, and the higher frequency
bands, compared to the PCEN and Multi-resolution (Figure~\ref{fig:ov_vanilla}).

\begin{figure}[H]
  \centering
  \includegraphics[width=0.92\linewidth]{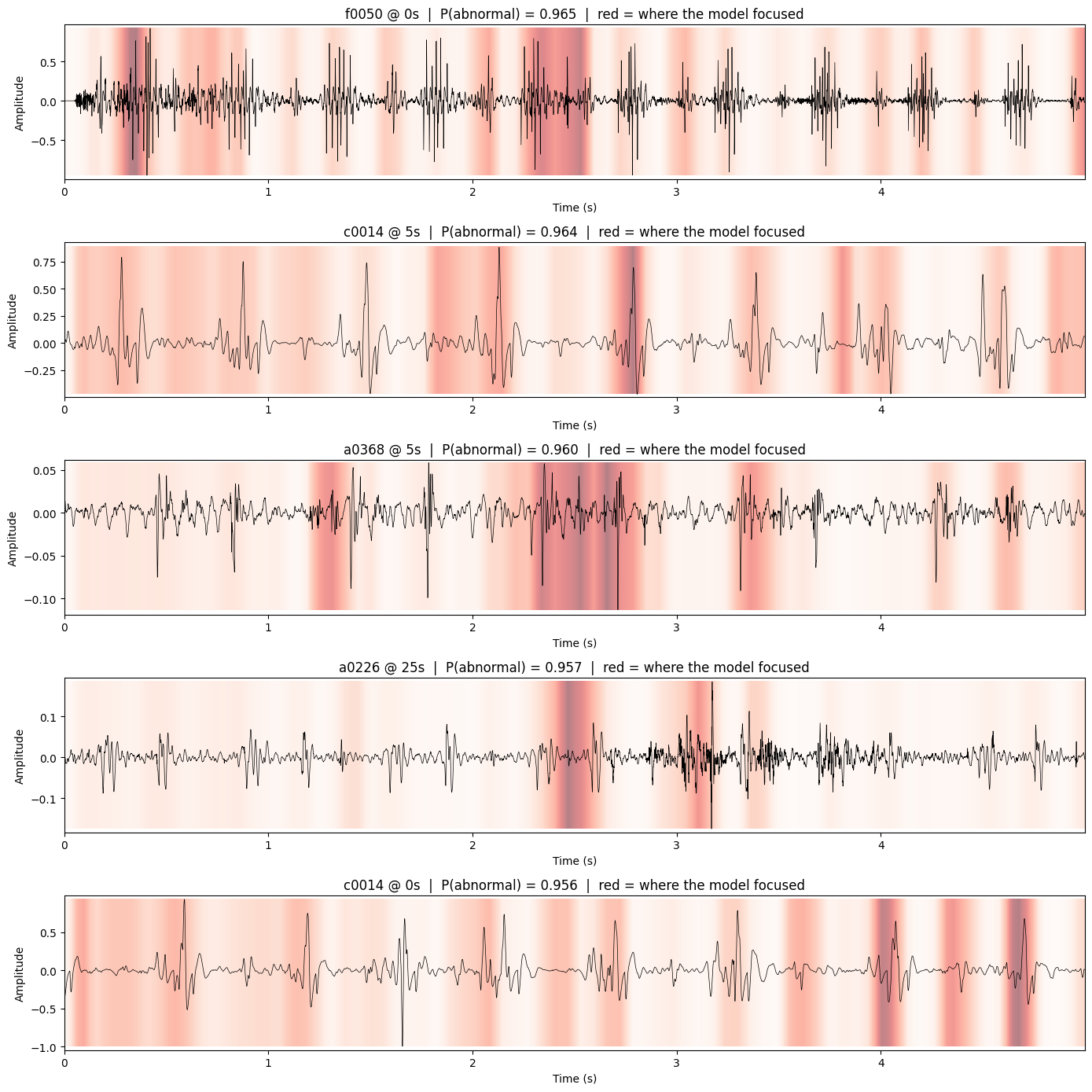}
  \caption{PCEN CNN. Grad-CAM attention overlaid on the waveform of a highly confident abnormal clip
  ($P(\text{abnormal}) = 0.967$). Attention sits on and just after the spikes --- the gap where murmurs
  appear.}
  \label{fig:ov_pcen}
\end{figure}

\paragraph{PCEN CNN.} The PCEN model mostly pays attention to the area in between S1 and S2. Both the
vanilla and PCEN models attend to the spikes (the heartbeat itself), but PCEN concentrates more on the
sound immediately after each spike, most likely focusing on murmurs (Figure~\ref{fig:ov_pcen}).

\paragraph{Multi-resolution CNN.} The multi-resolution model is not focused on specific areas. It
gets teh majortiy of the heartbeat, which stops overrlieance on any specific area of the waveform.

\begin{figure}[H]
  \centering
  \includegraphics[width=0.92\linewidth]{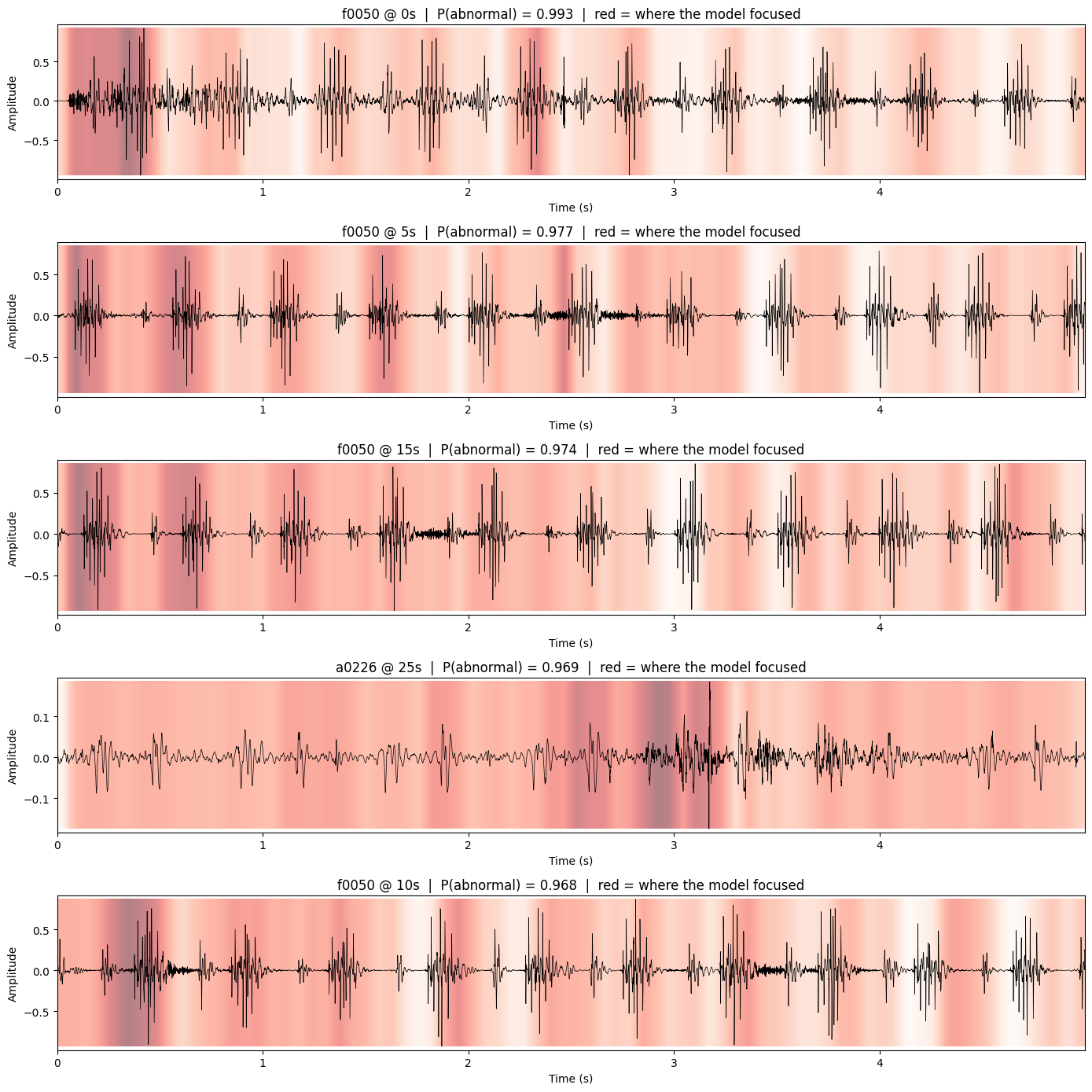}
  \caption{Multi-resolution CNN. Grad-CAM attention overlaid on the waveform of a confident abnormal
  clip ($P(\text{abnormal}) = 0.977$). Attention is spread broadly across the recording rather than
  pinned to a few beats.}
  \label{fig:ov_multi}
\end{figure}

Narrow-window spectrograms give great time resolution but poor frequency resolution, while wider-window
spectrograms are the opposite. Narrow windows detect arrhythmia (timing issues) better, while wider windows
detect murmurs, which are more relaetd to frequency. Because both sharp spikes and murmurs matter, the multi-resolution stack
attends to almost all regions (Figure~\ref{fig:ov_multi}).

\subsection{Answer to Question 3}
Accross all three of the front-ends, the biggest indicators of poor heart health
wre the timings adn the frequenc spikes of the S1 and S2 of the hearbeat. For PCEN and the vanilla log-mel
the attentio is more focused on teh S1 and S2 areas, while the multi-resolution front-end is more
spread out accross the entire signal.These observations would still require consultation with medical professionals to
verify, but a system like this could help physicians identify specific stress points in a recording. The
natural next goal would be to move beyond normal-versus-abnormal and identify exactly which condition a
given recording reflects

\section{Interpretation of Findings}
The results support a clear interpretation. Because the architecture,
epochs, and batches were identical across all three runs, the measured differences
 are attributable to thefront-end alone. That the two adaptive or richer representations (PCEN and multi-resolution log-mel)
both beat the static log-mel baseline --- and do so in the \emph{same} way, by reducing false alarms on
normal recordings --- is consistent with the intuition that motivated them. Static log compression
treats every frequency bin identically, which can let common, loud background structure dominate the
representation; PCEN's per-channel normalization instead emphasizes departures from each channel's
own baseline, and the multi-resolution stack gives the CNN several views of the same event. Both
make the abnormal-versus-normal boundary slightly easier to draw without changing how aggressively
the model pursues abnormal cases.

It is equally important that the gains are \emph{modest}, but not negligible. The vanilla log-mel 
baseline is already a strong detector, and the enhanced front-ends improve modified accuracy by 
roughly half a percent.
The right conclusion is therefore not that log-mel is
``bad,'' but that for this dataset and this compact CNN, the adaptive representations are a small, low-risk
improvement --- and the multi-resolution variant pays for its top score with a $3\times$ heavier
front-end.

The very high sensitivity across all models, paired with lower specificity, is a direct result of
the inverse-frequency class weighting. In a screening context this is a sensible default: a false negative
(missing an abnormal recording) is generally more costly than a false positive (flagging a normal
recording for an unnecessary second look), so a classifier that errs toward catching abnormal cases is
behaving the way a screening tool should. The modified-accuracy metric is well suited to this setting
precisely because it refuses to let a model coast on the normal majority; it forces sensitivity and
specificity to be balanced before it rewards the model.

Finally, the Grad-CAM evidence strengthens confidence in the whole pipeline. A model can achieve high
accuracy for the wrong reasons --- latching onto a recording-specific artifact or a dataset shortcut --- and
on an imbalanced dataset that risk is real. The fact that attention concentrates in the low-frequency
S1/S2 band, and lands on individual beats in the highest-confidence abnormal clips, indicates that the
network is learning a stable acoustic pattern tied to cardiac function rather than exploiting noise. This
is the kind of model-to-problem fit the project set out to demonstrate.

A useful way to frame these results is in terms of \emph{where the useful inductive bias enters the
pipeline}. With the architecture, optimizer, and training budget all fixed, the only thing that changes
between conditions is how much task-relevant structure the input representation exposes \emph{before}
any weights are learned. PCEN and the multi-resolution stack both inject domain-appropriate bias at the
front end: PCEN's per-channel automatic gain control suppresses slowly varying background energy so that
transient, locally-loud events (the murmurs and clicks that distinguish abnormal recordings) survive into
the representation, while the multi-resolution stack supplies several time--frequency trade-offs at once.
Because the CNN no longer has to discover that structure on its own from a fairly small dataset, the same
network settles on a slightly better decision boundary. This is the classic lesson that, in
low-to-moderate data regimes, the quality of the representation can contribute as much as the capacity of
the model.

The multi-resolution result in particular has a concrete signal-processing justification rooted in the
time--frequency uncertainty principle. A single spectrogram cannot be sharp in both time and frequency at
once: a short analysis window localizes events in time but blurs them in frequency, while a long window
does the opposite. Heart-sound abnormality lives on \emph{both} axes --- arrhythmia is largely a timing
phenomenon (favoring short windows), while murmurs are largely a spectral phenomenon (favoring long
windows) --- so no single window length is optimal. Stacking three window sizes as input channels lets the
convolutional filters draw on whichever resolution is most informative for a given pattern, which is
exactly why the multi-resolution front-end posts the best sensitivity and shows the broadest Grad-CAM
attention.

The Question~2 experiment sharpens this reading. Shrinking the network reduces its capacity to
\emph{compensate} for a weak input representation, so the quality of the front-end should matter more, not
less --- and that is what we see. The gap between the plain log-mel baseline and the two enhanced
front-ends widened in the smaller network, because the two-block CNN could no longer paper over an
un-normalized, single-resolution input. Read this way, part of the benefit of PCEN and multi-resolution is
that they \emph{pre-compute} structure a larger network might otherwise learn for itself, effectively
trading model capacity for representation quality.

Two caveats temper all of this. First, Grad-CAM is functioning here as a guard against shortcut learning:
on an imbalanced clinical dataset a classifier can reach high accuracy by keying on a recording-specific
artifact rather than cardiac physiology, and the fact that attention instead concentrates on the
low-frequency S1/S2 band is what gives us confidence the learned rule is physiologically grounded. Second,
the margins separating the front-ends are small relative to the run-to-run variance of training: re-running
the smaller-architecture experiment shifted modified accuracy by one to two points between seeds. The
front-end ranking should therefore be read as a tendency rather than an exact ordering, and a fully
rigorous comparison would average several random seeds or use cross-validation --- a natural extension once
more compute is available.

\section{Conclusion}
The front-end choice did not shift sensitivity much since all versions landed near $0.95$.
PCEN and the multi-resolution gad higher specificity and the modified accuracy score,
mostly by making fewer errors on normal cases, while log mel sat a bit lower at $0.91$.
Training one CNN across three audio representations under the same conditions, which let them
isolate what the input format itself contributed. Log mel served as the baseline, with PCEN and the
stacked multi-resolution version added in. Grad-CAM ended up focusing on low frequencies around the
heart sounds and murmurs.

Controlling the arhiceture and trainign conditions was crucial to making fair comparisons,
and the results show that the front-end does have a real impact on performance, but the differences are modest and not
game-changing. The vanilla log-mel is already a strong baseline, and the enhanced front-ends give a small but
consistent boost, especially in specificity. The multi-resolution front-end edges out the others on the official
PhysioNet metric, but it is not clear that the extra complexity is worth it given the small margin.
 The Grad-CAM analysis confirms that the models are learning to attend to meaningful acoustic patterns 
 rather than noise, which supports the validity of the whole pipeline.

To summarize, the project demonstrates that the choice of spectrogram front-end can have a real but modest
impact on abnormal heart-sound detection, and that adaptive representations like PCEN can help the model
focus on relevant audio patterns without changing everything.

\section{Improvements and Future Work}
Several extensions would improve upon this project. The most useful and obvious would be to test more of th
(abnormal) class, since the dataset is heavily imbalanced.

More computing power would also allow us to use and try larger spectrograms (for example, the $10$-second clips
that proved too slow to train here) and to
run more hyperparameter tuning. 
To improver interpretability, a more detailed analysis of the Grad-CAM
heatmaps so that we can identify the heart condition, not just whether a recording is abnormal or not.
This project was limited by lack of a reliably labeled dataset and by minimal medical knowledge
to actually identify the specific heart conditions and 
condition-specific features. Augmenting the spectrograms, like shifting the audio and slightly chaing it to
to create synthetic trainin data would also be a great new addition for furthur model training.
Lastly, since most of the heart-beats are recorded at resting state, it would be beneficail to train
the model of more variaous heart conditions, like extremely slow, and excersizing.

\section*{Dataset}
\noindent
PhysioNet/CinC 2016 Challenge --- Classification of Heart Sound Recordings:\\
\url{https://physionet.org/content/challenge-2016/}\\
Demo dataset:\\
\url{https://www.kaggle.com/datasets/kinguistics/heartbeat-sounds}

\section*{References Related to Research}
\begin{enumerate}[leftmargin=2em]
\item Liu, C., Springer, D., Li, Q., Moody, B., Juan, R.~A., Chorro, F.~J., Castells, F., Roig, J.~M.,
Silva, I., Johnson, A.~E.~W., Syed, Z., Schmidt, S.~E., Papadaniil, C.~D., Hadjileontiadis, L.,
Naseri, H., Moukadem, A., Dieterlen, A., Brandt, C., Tang, H., Samieinasab, M., Samieinasab, M.~R.,
Sameni, R., Mark, R.~G., \& Clifford, G.~D. (2016). An open-access database for the evaluation of
heart sound algorithms. \emph{Physiological Measurement}, 37(12), 2181--2213.
\item Potes, C., Parvaneh, S., Rahman, A., \& Conroy, B. (2016). Ensemble of Feature-based and
Deep Learning-based Classifiers for Detection of Abnormal Heart Sounds. \emph{Computing in
Cardiology}, 43, 621--624.
\item Rubin, J., Abreu, R., Ganguli, A., Nelaturi, S., Matei, I., \& Sricharan, K. (2017). Recognizing
Abnormal Heart Sounds Using Deep Learning. \emph{arXiv preprint} arXiv:1707.04642.
\end{enumerate}

\end{document}